\documentclass[aps,prx,reprint,noshowpacs,superscriptaddress,floatfix,letterpaper,longbibliography, notitlepage]{revtex4-1}
\usepackage{amsmath, amsfonts, amssymb, graphicx, hyperref, lineno,lipsum,setspace}
\usepackage{braket}
\usepackage{physics}
\usepackage{mathtools}
\usepackage[dvipsnames]{xcolor} 

\hypersetup{colorlinks=true, linkcolor=MidnightBlue,urlcolor=MidnightBlue,citecolor=MidnightBlue}
\usepackage[dvipsnames]{xcolor}

\begin{document}
\title{Geometry- and inertia-limited chaotic growth in classical many-body systems}

\author{Swetamber Das}
\email{swetamber.p@srmap.edu.in }

\affiliation{Department of Physics, SRM University--AP, Amaravati 522240, Andhra Pradesh, India }

\begin{abstract}
Chaotic instability in many-body systems is commonly quantified by the largest Lyapunov exponent, yet general constraints on its magnitude in classical interacting systems remain poorly understood. Here we establish explicit, Hamiltonian-specific upper bounds on the largest Lyapunov exponent for classical many-body systems with local interactions. These bounds arise from instantaneous stability constraints on the Hamiltonian flow and are expressed in terms of inertial scales and the curvature of the interaction potential. We show that they naturally separate into two qualitatively distinct classes: non-violable bounds, controlled by worst-case local curvature scales and inertia and insensitive to spatial structure, and ergodic ceilings, which retain spectral information and encode collective modes and finite-size effects under generic dynamical evolution. For a paradigmatic one-dimensional coupled-rotor chain (Josephson junction array), the ergodic ceiling admits a closed analytic form and produces a dynamically inaccessible region for sustained chaotic growth
in the Lyapunov exponent–energy plane, which we confirm numerically. In contrast to non-violable estimates, the ergodic ceiling yields a sharper constraint on chaotic growth by capturing collective suppression mechanisms absent at the level of local curvature alone. Remarkably, in the thermodynamic limit the ergodic ceiling asymptotically approaches an inertial ceiling that limits sustained Lyapunov growth, becoming independent of temperature and interaction strength. While classical systems do not admit universal chaos bounds, our results identify a broad class of natural Hamiltonian systems in which chaotic growth is inherently limited by inertia and interaction geometry, thereby setting a minimal microscopic timescale for long-time loss of memory of initial conditions.
\end{abstract}

\maketitle

{\textit{Introduction.}--- Classical many-body systems generically exhibit chaotic dynamics, manifested as
exponential sensitivity to initial conditions and typically quantified by Lyapunov
exponents. Chaos underlies mixing, equilibration, and the apparent loss of information
in deterministic systems, and plays a central role in statistical mechanics
\cite{Eckmann1985Ergodic,Gaspard1998Chaos}. Despite its ubiquity, a basic and surprisingly
unresolved question remains: are there intrinsic limits on how rapidly chaotic
instability itself can grow in interacting classical systems?

In quantum many-body systems, the growth of chaos is constrained by universal
bounds linking Lyapunov exponents to temperature and fundamental constants
\cite{Maldacena2016ChaosBound}. In contrast, classical Hamiltonian dynamics is
often assumed to allow arbitrarily fast chaotic growth once interactions become
sufficiently strong, with no analogue of a universal bound. This view has led to
the common perception that classical chaos is fundamentally unconstrained, aside
from phenomenological scaling relations for Lyapunov exponents based on
microscopic time scales, energy, or interaction strength \cite{Livi1986Lyapunov}.
However, classical dynamics are generated by a fixed Hamiltonian structure, which
imposes nontrivial constraints on instantaneous stability even in the absence of
dissipation or external driving.

Motivated by this observation, several works have explored possible restrictions
on classical chaos through conjectured scaling relations and asymptotic bounds
on the energy dependence of Lyapunov exponents, as well as geometric approaches
relating dynamical instability to curvature properties of the underlying
energy landscape~\cite{Casetti2000Geometry,Liu2021LyapunovTime,Watanabe2022}. While
these frameworks provide valuable insight into trends at high energy or strong
coupling, they do not address whether chaotic growth can be arbitrarily fast at
intermediate times, nor do they fully account for spatial organization and
collective modes in extended systems. Moreover, bounds based on long-time
averages, asymptotic scaling, or statistical assumptions need not constrain the
instantaneous dynamics and can, in principle, be transiently violated. From this
perspective, phenomenological scaling relations for Lyapunov exponents observed
in classical many-body systems can be viewed as describing how chaotic growth
explores the interior of a Hamiltonian-dependent envelope set by local stability
constraints. The bounds derived here place non-violable, instantaneous limits on this envelope,
independent of asymptotic scaling assumptions, thereby providing a framework within 
which observed energy-, size-, or temperature-dependent scaling
of Lyapunov exponents must reside.
\begin{figure*}[t]
\centering
\includegraphics[width=0.75\textwidth]{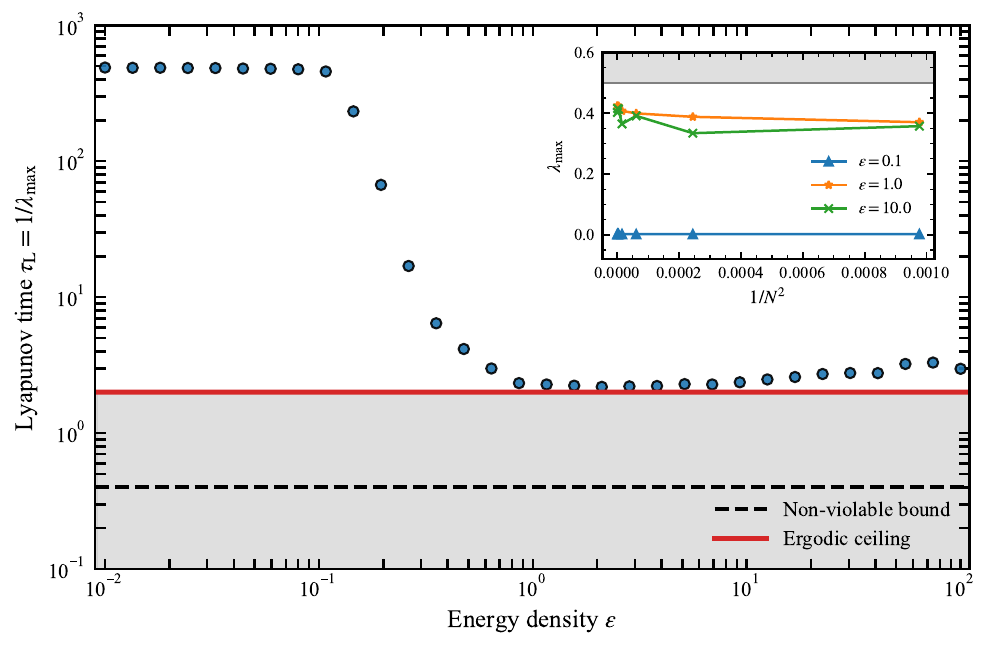}
\caption{
Lyapunov time $\tau_{\mathrm{L}} = 1/\lambda_{\max}$ as a function of energy density
$\varepsilon$ for the one-dimensional coupled rotor chain (Josephson junction array).
Blue filled circles show numerical values obtained from tangent-space dynamics, while
the solid and dashed curves denote the ergodic-ceiling and non-violable analytical
bounds, respectively. Parameters are $J=1$, $m=1$, and $N=2048$.
The shaded region marks a \emph{dynamically inaccessible} regime for sustained chaotic
growth, implied by the instantaneous stability constraint.
At high energy density, the ergodic ceiling approaches an inertial ceiling
$\tau_{\mathrm{L}} = 2m$, reflecting the diminishing role of collective modes, while the
non-violable bound remains a conservative, energy-independent constraint.
\emph{Inset:} Largest Lyapunov exponent $\lambda_{\max}$ as a function of $1/N^{2}$ at
fixed energy density, together with representative numerical values.
The $1/N^{2}$ scaling originates from the softest collective mode entering the ergodic
ceiling, illustrating how interaction geometry suppresses chaotic growth and controls
the approach to the inertial ceiling as $N\to\infty$. 
}

\label{fig:bound_plot}
\end{figure*}

In this Letter, we show that classical chaos is subject to explicit,
Hamiltonian-specific constraints once the instantaneous structure of the
equations of motion is taken into account. We derive Hamiltonian-specific
upper bounds on the largest Lyapunov exponent $\lambda_{\max}$, defined as the
long-time average of instantaneous stretching rates, for classical many-body
Hamiltonians with local interactions, expressed in terms of inertial scales and
the curvature of the interaction potential. A central feature of our approach
is that the bounds follow directly from instantaneous stability constraints,
determined by the largest eigenvalue of the \textit{symmetric} part of the
stability matrix, and require no statistical assumptions or long-time
averaging. As a result, the long-time growth of chaos is constrained by
instantaneous stability, enabling the identification of non-violable limits
on chaotic growth, as well as ergodic ceilings associated with finite-size
effects, interaction geometry, and an inertial ceiling that are inaccessible to
existing approaches. For clarity, we refer to the former as \emph{geometry-blind} bounds, since they depend only on local curvature and inertia, and to the latter as \emph{geometry-aware} ceilings, as they retain spectral information associated with collective modes and spatial structure.

 }

{\textit{Main result.}---We consider classical many-body systems with $n$ degrees of freedom, described by a natural Hamiltonian $H=\sum_{i=1}^{n}\frac{p_i^2}{2m}+V(\mathbf{q}),$ where $V(\mathbf{q})$ is a smooth interaction potential encoding interactions between degrees of freedom. Chaotic instability in such systems is quantified by $\lambda_{\max}$, which characterizes the exponential sensitivity of trajectories to infinitesimal perturbations in phase space and sets the fastest rate at which information about initial conditions is lost under deterministic time evolution. The growth of perturbations is governed by the linearized equations of motion, whose evolution is controlled by the stability matrix associated with the Hamiltonian flow.

By analyzing this linearized dynamics at the level of instantaneous stability, we derive rigorous upper bounds on $\lambda_{\max}$ that depend only on inertial scales and the local curvature of the interaction potential. Specifically, an infinitesimal perturbation $\delta\mathbf{x}(t)$ about a reference trajectory evolves as $d_t{\delta\mathbf{x}}=\mathbf{A}(t)\,\delta\mathbf{x}$, where $\mathbf{A}(t)$ is the stability matrix evaluated along the trajectory. The instantaneous growth or contraction of perturbations is controlled by the
symmetric part of $\mathbf{A}(t)$, whose eigenvalues define the local stretching
rates in phase space; the largest eigenvalue sets the maximal local
stretching rate $\sigma_{\max}(t)$~\cite{Das2025SpectralBounds}.  For Hamiltonians of the form considered here, the symmetric part of
$\mathbf{A}(t)$ is determined by the Hessian $\nabla^2 V(\mathbf{q}(t))$,
which therefore controls the instantaneous stretching rates. Because $\sigma_{\max}(t)$ is determined entirely by the local configuration of the system, it provides a pointwise constraint on chaotic growth, ensuring that instability is limited by local stability properties rather than long-time averages. This instantaneous foundation distinguishes our results from bounds based on long-time averages or asymptotic scaling arguments. 

Averaging the local stability constraint along a trajectory yields an upper bound
on $\lambda_{\max}$. Since the instantaneous stretching rate is pointwise bounded
by the symmetric part of the stability matrix, any such bound directly constrains
the largest Lyapunov exponent. This leads to the non-violable inequality~(see Supplementary Material, SM I):
\begin{equation}
\lambda_{\max}
\le
\frac{1}{2m}
+
\frac{1}{2}
\left\langle
\|\nabla^2 V(\mathbf{q})\|
\right\rangle ,
\end{equation}
where $\|\nabla^2 V\|$ denotes the operator norm of the Hessian, i.e., the magnitude
of its largest eigenvalue, which controls the maximal instantaneous stretching rate
generated by the symmetric part of the stability matrix. Angular brackets denote a time average along the trajectory; the bound remains non-violable because the instantaneous stretching rate is pointwise constrained at all times.
We work in units where the coordinates $q_i$ are
dimensionless, so that the interaction Hessian $\nabla^2 V$ sets an inverse--time--squared
scale (up to a factor of the mass $m$), while the mass $m$ itself has dimensions of time (see SM II).
The bound therefore reflects the interplay of inertia and local phase-space curvature,
while remaining blind to the spatial organization of the system and the structure of
its collective modes. Physically, the term $1/2m$ represents an inertial bottleneck:
finite inertia limits how rapidly phase-space trajectories can respond to forces,
placing an absolute constraint on the rate of chaotic growth. Since $\lambda_{\max}$
is the long-time average of the instantaneous stretching rate, any pointwise bound
on the latter directly implies a non-violable, instantaneous bound on $\lambda_{\max}$, even though
finite-time Lyapunov exponents fluctuate.

A tighter constraint is obtained by retaining the full spectral structure of the Hessian rather than its norm. This ergodic-ceiling refinement incorporates information about collective modes and long-wavelength excitations, yielding explicit finite-size corrections to chaotic growth. Importantly, the spectral radius entering the bound does not correspond to a fine-tuned static configuration, but instead encodes the strongest collective amplification channel permitted by the interaction geometry. As a result, the ergodic ceiling constrains the envelope of chaotic growth accessible under generic dynamical evolution and is strictly tighter than the non-violable estimate for any finite system size, as we show below.
}

{ \textit{Model.}—We first apply our bounds to a paradigmatic one-dimensional many-body
system: a chain of coupled rotors, which provides a classical description of a
one-dimensional Josephson junction array~\cite{Casetti2000Geometry,Mithun2019DynamicalGlass}.
For this model, the number of degrees of freedom introduced above coincides with the number of rotors, so that 
$n=N$. The system is described by the Hamiltonian:
\begin{equation}
\mathcal{H}(p,q)
=
\sum_{i=1}^N \frac{p_i^2}{2m}
+
\sum_{i=1}^N
J\bigl[1-\cos(q_{i+1}-q_i)\bigr],
\end{equation}
with periodic boundary conditions $q_{N+1}=q_1$ and $p_{N+1}=p_1$.
The variables $q_i$ represent angular degrees of freedom; in the Josephson-junction
interpretation they correspond to superconducting phase differences.
Equivalently, the same Hamiltonian describes a classical XY chain, where the kinetic
terms arise from grain charging energies and the cosine interactions encode
nearest-neighbor phase coupling.
This system constitutes a minimal example of an inertial many-body Hamiltonian with
bounded interactions, local coupling, and a well-defined interaction geometry.

The coupled-rotor chain is particularly well suited for illustrating our bounds for two reasons. First, the interaction potential has an analytically tractable form, allowing both the local curvature and the full spectral properties of the Hessian to be evaluated explicitly~\cite{Casetti2000Geometry}. As a result, the ergodic ceiling admits a closed analytic expression, enabling the identification of \textit{dynamically inaccessible regions}  in the Lyapunov exponent–energy plane. Second, the model exhibits a broad chaotic regime controlled by temperature and interaction strength, providing an ideal testbed for direct comparison between analytical bounds and numerically computed Lyapunov exponents across a wide range of parameters~\cite{Mithun2019DynamicalGlass,Skokos2009Delocalization}.

Apart from analytical convenience, the rotor chain captures physical ingredients common to many extended Hamiltonian systems, including inertia, nonlinearity, and collective modes. In particular, long-wavelength excitations of the phase variables play a central role in suppressing chaotic growth at finite system size, directly illustrating the geometry-induced constraints encoded in our bounds. We emphasize that this suppression mechanism is distinct from slow transport phenomena such as Arnold diffusion or phase-space bottlenecks near integrable limits. Instead, the bounds derived here constrain the instantaneous local stretching rate and therefore apply deep in the chaotic regime, independent of long-time transport or equilibration processes.

In thermal equilibrium, the non-violable bound is controlled by the operator norm
of the interaction Hessian, which reflects the largest instantaneous local curvature
accessible to the Hamiltonian dynamics. For the rotor chain, the boundedness of the
interaction implies the uniform estimate $\|\nabla^2 V\| \le 4J$, independent of temperature, 
system size, or spatial structure. Substituting this
estimate into the general bound yields the non-violable constraint~(see SM I):
\begin{equation}
\lambda_{\max}
\le
\frac{1}{2m}
+
2J ,
\end{equation}
which provides a conservative, Hamiltonian-level upper bound on chaotic growth that
depends only on inertial and local interaction scales. While necessarily loose, this
bound applies to arbitrary trajectories and highlights the limitations of local
curvature-based estimates in constraining chaos without incorporating collective
geometric information.

A sharper bound is obtained by retaining the full spectral information of the
interaction Hessian. For systems with nearest-neighbor cosine interactions, 
the ergodic ceiling incorporates collective modes encoded in the
discrete Laplacian spectrum, yielding (see SM III):
\begin{equation}
\lambda_{\max}
\lesssim
\frac{1}{2m}
\left[
1
-
\frac{4\pi^2 J}{N^2}
\frac{I_1(\beta J)}{I_0(\beta J)}
\right].
\end{equation}
where $I_n$ denotes the modified Bessel function of the first kind of order $n$.
This expression applies directly to the rotor chain, where the inertial ceiling
$1/2m$ originates from the kinetic term, while the finite-size correction reflects
geometry-induced suppression of chaotic growth by the softest collective mode.
In one dimension, this correction scales as $1/N^2$, corresponding to the
longest-wavelength  mode. More generally, because the correction
is controlled by the smallest nonzero eigenvalue of the interaction Laplacian,
its scaling is geometric in origin and depends on spatial dimensionality,
behaving as $N^{-2/d}$ in $d$ dimensions. In the thermodynamic limit $N\to\infty$ at fixed energy density, 
the geometry-dependent term vanishes and the bound saturates to the inertial ceiling, becoming independent
of temperature and interaction strength.

Beyond furnishing absolute ceilings on chaotic growth, the bounds introduced here
clarify the distinct mechanisms by which instability is constrained in classical
many-body systems. The non-violable bound reflects the maximal instantaneous
instability permitted by inertia and local curvature alone and is necessarily
conservative. In contrast, the ergodic ceiling encodes additional suppression
arising from spatial organization, collective modes, and finite-size effects.
The fact that only the ergodic ceiling provides a sharp, energy- and
size-dependent constraint highlights the essential role of interaction geometry
in regulating chaotic growth. The degree to which the largest Lyapunov exponent
approaches this ergodic ceiling therefore serves as a quantitative indicator
of how effectively collective dynamics suppress local instability under generic dynamical evolution, without assuming  ergodicity of the dynamics. Because these bounds
are constructed from instantaneous stability considerations, they constrain
chaotic growth at all times and provide insight into the structure of the dynamics
without requiring explicit computation of asymptotic scaling laws.\\

}
{\textit{Numerical implementation.}--- For Fig.~\ref{fig:bound_plot}, the largest
Lyapunov exponent $\lambda_{\max}$ is computed using standard tangent-space methods
by integrating the equations of motion together with their linearized dynamics
\cite{Benettin1980Lyapunov}. Instantaneous stretching rates are resolved by periodic
renormalization of the tangent vectors, and $\lambda_{\max}$ is obtained from the
accumulated growth of their norm. Time evolution is performed using the second-order
symplectic integrator SBAB$_2$, ensuring excellent long-time energy conservation
\cite{LaskarRobutel2001,Skokos2009Delocalization}. Initial conditions are sampled
from the microcanonical ensemble at fixed energy density. The main panel of
Fig.~\ref{fig:bound_plot} shows the Lyapunov time $\tau_{\mathrm{L}}=1/\lambda_{\max}$
versus energy density together with the non-violable bound and the ergodic ceiling,
while the inset illustrates the finite-size scaling of $\lambda_{\max}$ with the
$1/N^2$ dependence arising from the softest collective mode entering the ergodic
ceiling. Further numerical details are provided in SM IV.}

{\textit{Discussion and Outlook.}--- The results presented here establish that chaotic growth in
classical many-body systems is subject to intrinsic, Hamiltonian-specific constraints once the
microscopic dynamics is fixed. Although classical Hamiltonian systems do not admit
a universal chaos bound analogous to those in quantum mechanics, we show that
inertia and interaction geometry impose explicit upper limits on the largest
Lyapunov exponent $\lambda_{\max}$. These limits arise from instantaneous stability
constraints on the Hamiltonian flow and therefore apply at the level of instantaneous stability, 
independent of assumptions about ergodicity, mixing, or
equilibration. A central conceptual feature of our approach is the clear
separation between such deterministic, instantaneous constraints and properties of
long-time dynamics. The \emph{non-violable bound} follows directly from local curvature
and inertia and is uniform across trajectories, while the \emph{ergodic ceiling}
depends on spectral properties of the interaction network and captures collective
and finite-size effects. Consequently, the latter constrains the envelope of chaotic
growth accessible under typical dynamical evolution, while the former
provides a trajectory-level constraint that cannot be transiently violated. This
distinction delineates the domains of validity of the different bounds and
highlights the combined role of inertia and interaction geometry in preventing
arbitrarily rapid chaotic growth in deterministic many-body dynamics.

It is instructive to relate these results to semiclassical notions of
quantum--classical correspondence and information scrambling. In semiclassical
systems, the Ehrenfest time $t_E$ characterizes the duration over which quantum and
classical dynamics remain approximately aligned and is typically controlled by the
maximal Lyapunov exponent as
$t_E \sim \lambda_{\max}^{-1}\ln(S/\hbar)$
\cite{BermanZaslavsky1978,LarkinOvchinnikov1969,AleinerLarkin1996}, reflecting the
exponential amplification of initially small quantum uncertainties by classical
chaos. From this perspective, the bounds derived here indirectly constrain the
Ehrenfest time by placing Hamiltonian-specific upper limits on $\lambda_{\max}$. In
particular, the existence of an inertial ceiling on chaotic growth implies a
corresponding lower bound on $t_E$, which becomes independent of interaction
strength or temperature in the thermodynamic limit. Closely related consequences
arise for out-of-time-order correlators (OTOCs)
\cite{Khemani2018VelocityDependent,Bilitewski2018, Beims2022, Haneder2025relationclassicalquantumlyapunov}, 
which diagnose operator growth and information scrambling in quantum many-body systems.
In the semiclassical regime, early-time OTOC growth is governed by classical
phase-space instability, with a rate set by maximal classical stretching. By
bounding this stretching rate, inertia and interaction geometry therefore limit
the fastest possible growth of quantum operators inherited from the underlying
classical dynamics.

Beyond their semiclassical implications, the bounds derived here may be viewed as
imposing a dynamical speed limit on chaotic growth in classical Hamiltonian systems
\cite{Ohzeki2018,delCampo2018,Das2023SpeedLimitsChaos}, complementing earlier proposals
of chaos bounds in specific semiclassical settings \cite{Morita2020Bound}. By
constraining instantaneous stretching rates in phase space, they identify a minimal
microscopic timescale for the loss of memory of initial conditions once the
Hamiltonian is specified. In this sense, our results are complementary to the
quantum chaos bounds proposed by Maldacena--Shenker--Stanford
\cite{Maldacena2016ChaosBound}, which limit the growth of OTOCs at fixed temperature.
While the mechanisms and scope are fundamentally different, and no universality is
implied classically, the present framework provides explicit, constructive bounds
tied directly to inertia and interaction geometry. This Hamiltonian-level specificity
makes the bounds directly applicable to concrete models, offering a practical route
to diagnosing and constraining chaotic dynamics.

While our analysis focuses on classical systems with local interactions, the
underlying framework admits several natural extensions. In higher dimensions, the
modified spectrum of collective modes suggests that geometry-induced suppression
of chaos acquires a distinct dimensional dependence, while disordered lattices and
networks with nontrivial connectivity provide promising settings in which spatial
heterogeneity may lead to nonuniform bounds on chaotic growth. Although long-range
interactions require additional care, the instantaneous nature of the bound
suggests that meaningful constraints may persist even in such settings. 
Lyapunov exponents are also closely related to mixing and equilibration; however, the bounds
derived here do not directly determine thermalization times, which can be governed
by distinct dynamical mechanisms and, in specific models, may even coincide with
instability timescales associated with confined or deconfined chaos
\cite{Malishava2022LyapunovSpectrum,Kim2025ConfinedChaos}. Instead, they establish a
minimal microscopic timescale for the onset of irreversibility, beyond which
equilibration, transport, and relaxation processes must operate. An important open
question is how such instantaneous chaos bounds constrain nonequilibrium
phenomena, including transport, entropy production, and driven dynamics. Taken
together, our results suggest that instantaneous stability provides a unifying
principle for bounding dynamical complexity in interacting many-body systems,
complementing approaches based on asymptotic scaling and hydrodynamic descriptions.
}

\textit{Acknowledgments.}—
The author acknowledges the use of the High Performance Computing Cluster
at SRM University–AP.

\bibliography{reference_chaos}

\clearpage
\onecolumngrid

\begin{center}
\textbf{\large Supplemental Material for}\\[6pt]
\textbf{\large Geometry- and inertia-limited chaotic growth in classical many-body systems}\\[10pt]
Swetamber Das\\
Department of Physics, SRM University--AP, Amaravati 522240, Andhra Pradesh, India\\
Email: swetamber.p@srmap.edu.in
\end{center}

\vspace{10pt}

\setcounter{equation}{0}
\setcounter{figure}{0}
\setcounter{section}{0}

\renewcommand{\thesection}{SM~\Roman{section}}
\renewcommand{\theequation}{S\arabic{equation}}
\renewcommand{\thefigure}{S\arabic{figure}}

\section{Non-violable (Geometry-Blind) bound from operator norms}

\subsection*{Linearized dynamics and stability matrix}

We consider a natural Hamiltonian system of the form:
\begin{equation}
H(\mathbf{q},\mathbf{p}) = \sum_{i=1}^n \frac{p_i^2}{2m} + V(\mathbf{q}),
\end{equation}
with equations of motion:
\begin{equation}
\dot{\mathbf{q}} = \frac{\mathbf{p}}{m},
\qquad
\dot{\mathbf{p}} = -\nabla V(\mathbf{q}).
\end{equation}

Linearizing the dynamics about a reference trajectory
$(\mathbf{q}(t),\mathbf{p}(t))$ yields the tangent-space evolution:
\begin{equation}
\frac{d}{dt}
\begin{pmatrix}
\delta \mathbf{q} \\
\delta \mathbf{p}
\end{pmatrix}
=
\mathbf{A}(\mathbf{q}(t))
\begin{pmatrix}
\delta \mathbf{q} \\
\delta \mathbf{p}
\end{pmatrix},
\end{equation}
where the stability matrix $\mathbf{A}$ is given by:
\begin{equation}
\mathbf{A}(\mathbf{q}) =
\begin{pmatrix}
\mathbf{0} & m^{-1}\mathbf{I} \\
-\nabla^2 V(\mathbf{q}) & \mathbf{0}
\end{pmatrix}.
\end{equation}
Here $\nabla^2 V$ denotes the Hessian of the potential, which is symmetric for smooth interaction potentials.

We denote the number of degrees of freedom by $n$ in the general discussion and by
$N$ when specializing to the rotor chain; the two should be identified in the latter case. Throughout the Supplementary Material, angular brackets $\langle\cdot\rangle$ denote ensemble averages, unless stated otherwise. We use the terms ``geometry-blind'' and
``geometry-aware'' as descriptive qualifiers for the non-violable bounds and ergodic
ceilings, respectively, indicating whether spatial and spectral information enters
the corresponding constraint.

\subsection*{Symmetric part and instantaneous growth rate}

The instantaneous exponential growth rate of tangent vectors is governed by
the symmetric part of the stability matrix:
\begin{equation}
\mathbf{A}_+ \equiv \frac{1}{2}\left(\mathbf{A} + \mathbf{A}^T\right).
\end{equation}
Explicitly, one finds
\begin{equation}
\mathbf{A}_+ =
\frac{1}{2}
\begin{pmatrix}
\mathbf{0} & m^{-1}\mathbf{I} - \nabla^2 V(\mathbf{q}) \\
m^{-1}\mathbf{I} - \nabla^2 V(\mathbf{q}) & \mathbf{0}
\end{pmatrix}.
\end{equation}

For any differentiable norm on phase space, the instantaneous growth rate
$r(t)$ of tangent vectors satisfies:
\begin{equation}
r(t)
\le
\lambda_{\max}\!\left(\mathbf{A}_+(t)\right),
\end{equation}
where $\lambda_{\max}$ denotes the largest eigenvalue of the symmetric matrix $\mathbf{A}_+$.

\subsection*{Operator-norm (non-violable) bound}

To obtain a non-violable bound, we decompose $\mathbf{A}_+$ as:
\begin{equation}
\mathbf{A}_+ = \frac{1}{2}\left(\mathbf{M} + \mathbf{N}\right),
\end{equation}
with
\begin{equation}
\mathbf{M} =
\begin{pmatrix}
\mathbf{0} & m^{-1}\mathbf{I} \\
m^{-1}\mathbf{I} & \mathbf{0}
\end{pmatrix},
\qquad
\mathbf{N} =
\begin{pmatrix}
\mathbf{0} & -\nabla^2 V(\mathbf{q}) \\
-\nabla^2 V(\mathbf{q}) & \mathbf{0}
\end{pmatrix}.
\end{equation}

Using subadditivity of the induced operator norm:
\begin{equation}
\lambda_{\max}(\mathbf{A}_+)
\le
\frac{1}{2}
\left(
\|\mathbf{M}\| + \|\mathbf{N}\|
\right).
\end{equation}
Here $\|\cdot\|$ denotes the operator norm induced by the Euclidean vector norm.

Since $\mathbf{M}$ and $\mathbf{N}$ are symmetric block matrices, their operator norms are given by:
\begin{equation}
\|\mathbf{M}\| = \frac{1}{m},
\qquad
\|\mathbf{N}\| = \|\nabla^2 V(\mathbf{q})\|.
\end{equation}
Consequently, the instantaneous growth rate satisfies the bound:
\begin{equation}
r(t)
\le
\frac{1}{2m}
+
\frac{1}{2}\,
\|\nabla^2 V(\mathbf{q}(t))\|.
\end{equation}

This bound depends only on the local curvature scale of the potential and is insensitive to the spatial structure or spectral properties of $\nabla^2 V$, and therefore constitutes a \emph{non-violable} constraint on instantaneous chaotic growth.

\subsection*{Bound on the largest Lyapunov exponent}

The largest Lyapunov exponent is defined as the long-time average of the
instantaneous exponential growth rate:
\begin{equation}
\lambda_{\max}
=
\lim_{t\to\infty}\frac{1}{t}
\int_0^t r(t')\,dt' .
\end{equation}
Using the instantaneous bound derived above:
\begin{equation}
r(t)
\le
\frac{1}{2m}
+
\frac{1}{2}
\|\nabla^2 V(\mathbf{q}(t))\|.
\end{equation}
Since the instantaneous growth rate is pointwise bounded at all times, the long-time average of $r(t)$ is directly constrained, independent of ergodicity or mixing properties. This yields the non-violable bound:
\begin{equation}
\lambda_{\max}
\le
\frac{1}{2m}
+
\frac{1}{2}
\left\langle
\|\nabla^2 V(\mathbf{q})\|
\right\rangle .
\end{equation}

This bound depends only on the local curvature scale of the potential and does not retain information about the spatial structure or spectral properties of the Hessian, hence the designation \emph{non-violable}.

\subsection*{Specialization to the one-dimensional rotor chain}

We now specialize the non-violable bound to the one-dimensional rotor
(Josephson junction) chain with Hamiltonian:
\begin{equation}
H = \sum_{i=1}^N \frac{p_i^2}{2m}
+ J\sum_{i=1}^N
\left[
1-\cos(q_{i+1}-q_i)
\right],
\end{equation}
subject to periodic boundary conditions.

For this model, the Hessian of the potential has the tridiagonal Laplacian form:
\begin{equation}
(\nabla^2 V)_{ij}
=
J\Big[
\big(\cos(q_{i+1}-q_i)+\cos(q_i-q_{i-1})\big)\delta_{ij}
-
\cos(q_{i+1}-q_i)\delta_{i,j-1}
-
\cos(q_i-q_{i-1})\delta_{i,j+1}
\Big].
\end{equation}
The operator norm of such a matrix can be bounded using standard row-sum (Gershgorin)
estimates. Applying the triangle inequality yields the instantaneous bound:
\begin{equation}
\|\nabla^2 V\|
\le
2J \max_i\!\left(
|\cos(q_{i+1}-q_i)| + |\cos(q_i-q_{i-1})|
\right)
\le
4J,
\end{equation}
which holds uniformly for all configurations, independent of temperature, system
size, or spatial correlations.

Substituting this estimate into the general non-violable bound for the instantaneous
growth rate gives:
\begin{equation}
r(t)
\le
\frac{1}{2m}
+
2J,
\end{equation}
which holds uniformly for all times and trajectories. Integrating this inequality
immediately yields the corresponding bound on the largest Lyapunov exponent:
\begin{equation}
\lambda_{\max}
\le
\frac{1}{2m}
+
2J.
\end{equation}

This non-violable bound depends only on inertial and local curvature scales and
does not retain information about the spatial structure or spectral properties of
the Hessian. In particular, it is insensitive to collective modes, thermal
fluctuations, and finite-size effects, in contrast to the ergodic ceiling
derived in the following subsection.

\section{Hamiltonian time units and dimensional consistency}

The stability matrix governing tangent-space evolution defines instantaneous
stretching with respect to the natural time parametrization of the Hamiltonian
flow. Throughout this work, Lyapunov exponents are therefore reported in the
dimensionless units associated with this intrinsic Hamiltonian timescale.

For the one-dimensional rotor (Josephson junction) chain, the corresponding
microscopic time scale is:
\begin{equation}
t_0 \sim \sqrt{\frac{m}{J}},
\end{equation}
which follows from balancing inertial and interaction terms in the equations of
motion. Physical Lyapunov exponents are obtained by converting from Hamiltonian
time to laboratory time according to:
\begin{equation}
\lambda_{\max}^{\mathrm{phys}} = t_0^{-1}\,\lambda_{\max}.
\end{equation}

All analytical bounds and numerical results presented in the main text and
Supplemental Material are formulated in these natural Hamiltonian units, ensuring
dimensional consistency across models and parameter regimes.

\section{Ergodic (Geometry-Aware) ceiling from collective modes and spectral structure}

\subsection*{Spectral structure of the symmetric stability matrix}

The symmetric part of the stability matrix introduced in the previous section is:
\begin{equation}
\mathbf{A}_+
=
\frac{1}{2}
\begin{pmatrix}
\mathbf{0} & m^{-1}\mathbf{I}-\nabla^2 V \\
m^{-1}\mathbf{I}-\nabla^2 V & \mathbf{0}
\end{pmatrix}.
\end{equation}
Owing to its block off-diagonal structure, the eigenvalues of $\mathbf{A}_+$ occur in pairs $\pm \mu/2$, where $\mu$ are the eigenvalues of the symmetric matrix:
\begin{equation}
\mathbf{B}
=
m^{-1}\mathbf{I}
-
\nabla^2 V .
\end{equation}
Consequently, the instantaneous growth rate satisfies:
\begin{align}
r(t)
&\le
\frac{1}{2}\,
\lambda_{\max}(\mathbf{B}) \nonumber \\
&=
\frac{1}{2}
\max\Bigl(
m^{-1}
-
\kappa_{\min}^{\text{(nonzero)}}(t),\ 
m^{-1}
-
\kappa_{\max}(t)
\Bigr),
\end{align}
where $\kappa_{\min}^{\text{(nonzero)}}$ and $\kappa_{\max}$ denote the smallest nonzero and largest eigenvalues of the Hessian $\nabla^2 V$, respectively, evaluated along the trajectory. Here the Hessian eigenvalues $\kappa_k$ are indexed by $k=0,\ldots,N-1$, which label the normal modes of the chain.
For the rotor chain with $J>0$, the largest curvature mode is bounded and does not control the maximal stretching rate, leading to the simpler bound:
\begin{equation}
r(t) \le \frac{1}{2} \Bigl( m^{-1} - \kappa_{\min}^{\text{(nonzero)}}(t) \Bigr).
\end{equation}
This shows that instantaneous chaotic growth is controlled by the softest {\it nonzero} curvature mode of the interaction potential. When combined with a statistical characterization of this soft mode along typical trajectories, this observation leads to an ergodic ceiling rather than a trajectory-level constraint.

\subsection*{Soft-mode control of chaotic growth in the rotor chain}

For the one-dimensional rotor (Josephson junction) chain, the Hessian of the potential is given by:
\begin{equation}
(\nabla^2 V)_{ij}
=
J \cos(q_i-q_{i-1})
\Bigl(
2\delta_{ij}
-
\delta_{i,j+1}
-
\delta_{i,j-1}
\Bigr),
\end{equation}
with periodic boundary conditions. 

In thermal equilibrium, the stiffness $J\cos(q_i-q_{i-1})$ fluctuates about its mean value:
\begin{equation}
\langle J\cos(q_i-q_{i-1}) \rangle
=
J\,\frac{I_1(\beta J)}{I_0(\beta J)}.
\end{equation}
Replacing the instantaneous stiffness by this thermal average yields an effective uniform-stiffness discrete Laplacian, whose eigenvalues are known exactly:
\begin{align}
\kappa_k
&=
2J\frac{I_1(\beta J)}{I_0(\beta J)}
\Bigl[
1
-
\cos\!\Bigl(\frac{2\pi k}{N}\Bigr)
\Bigr], \qquad
k=0,1,\dots,N-1 .
\end{align}
The zero mode $k=0$ ($\kappa_0=0$) reflects the global rotational invariance of the chain, while the smallest nonzero eigenvalue corresponds to the longest-wavelength collective mode $k=1$:
\begin{align}
\kappa_1
&=
4J\frac{I_1(\beta J)}{I_0(\beta J)}\,
\sin^2\!\Bigl(\frac{\pi}{N}\Bigr).
\end{align}
For $N\gg1$, $\sin^2(\pi/N) \approx (\pi/N)^2$, giving:
\begin{align}
\kappa_1
&\simeq
\frac{4\pi^2 J}{N^2}
\frac{I_1(\beta J)}{I_0(\beta J)} .
\end{align}
Substituting this expression for the smallest nonzero eigenvalue into the instantaneous bound yields:
\begin{align}
r(t) \lesssim \frac{1}{2m}
\Bigl[
1
-
\frac{4\pi^2 J}{N^2}
\frac{I_1(\beta J)}{I_0(\beta J)}
\Bigr].
\end{align}

\subsection*{Final ergodic ceiling}

Assuming typical dynamical evolution, the long-time average of the
instantaneous growth rate $r(t)$ gives the largest Lyapunov exponent. The only approximation entering this step is the replacement of the fluctuating stiffness by its thermal average; no additional assumptions about the form of the dynamics are made.
Using the mean-stiffness approximation described above, we obtain the ergodic ceiling:
\begin{align}
\;
\lambda_{\max}
\lesssim
\frac{1}{2m}
\Bigl[
1
-
\frac{4\pi^2 J}{N^2}
\frac{I_1(\beta J)}{I_0(\beta J)}
\Bigr]
\;.
\end{align}
The symbol $\lesssim$ indicates that the inequality follows from replacing the fluctuating stiffness by its thermal average; direct numerical integration confirms that the bound holds across all studied parameter regimes. 

This result explicitly encodes spatial structure and finite-size effects through the softest collective mode of the system. In contrast to the non-violable bound, the ergodic ceiling exhibits a finite-size suppression of chaotic growth that scales as $N^{-2}$ and vanishes in the thermodynamic limit, leaving the inertial ceiling $\lambda_{\max} \le 1/(2m)$.

\section{Numerical Implementation and Computational Details}

We compute the largest Lyapunov exponent of the one-dimensional rotor (Josephson junction) chain using standard tangent-space methods combined with symplectic time integration. This section provides full details of the numerical protocol used for all results reported in the main text and in the finite-size scaling analysis.

\subsection*{Model}

We consider a one-dimensional chain of $N$ coupled rotors described by the Hamiltonian:
\begin{equation}
H = \sum_{i=1}^{N} \frac{p_i^2}{2m}
+ J \sum_{i=1}^{N} \left[ 1 - \cos(q_{i+1} - q_i) \right],
\end{equation}
with periodic boundary conditions. Throughout the simulations, we fix the coupling strength $J=1$ and mass $m=1$, which sets the natural energy and time scales.

\subsection*{Initial conditions and energy control}

Initial rotor angles are chosen uniformly as:
\begin{equation}
q_i(0) = 0 \qquad \forall\, i .
\end{equation}
Initial momenta are drawn independently from a Maxwell distribution with small width and subsequently adjusted to remove the center-of-mass motion:
\begin{equation}
p_i \rightarrow p_i - \frac{1}{N}\sum_{j=1}^N p_j .
\end{equation}
To impose a prescribed energy density $\varepsilon = E/N$, the momenta are rescaled according to:
\begin{equation}
p_i \rightarrow p_i
\sqrt{\frac{2N\varepsilon}{\sum_j p_j^2}} .
\end{equation}
After rescaling, the energy density is verified numerically to machine precision before time integration begins.

\subsection*{Time integration}

Time evolution is performed using the second-order symplectic integrator SBAB$_2$,
which is well suited for Hamiltonian systems with separable kinetic and potential
terms and provides excellent long-time energy conservation. The same integrator is
used for both the reference trajectory and the tangent-space dynamics.

A fixed time step
\begin{equation}
\Delta t = 0.01
\end{equation}
is used throughout. The total integration time is
\begin{equation}
T = 4000,
\end{equation}
corresponding to $4\times10^{5}$ integration steps. An initial transient corresponding
to a fraction
\begin{equation}
f_{\mathrm{tr}} = 0.05
\end{equation}
of the total integration time is discarded to eliminate dependence on the initial
condition.

Energy conservation is monitored throughout the Lyapunov measurement phase. For all
system sizes and energy densities considered, the maximum relative energy drift
remains below $10^{-8}$. We verified timestep convergence by repeating representative simulations with a
smaller timestep $\Delta t=0.005$, finding no statistically significant change in
the extracted Lyapunov exponent within numerical resolution.

\subsection*{Tangent-space dynamics and Lyapunov exponent}

The largest Lyapunov exponent is computed by evolving an infinitesimal perturbation vector in tangent space alongside the reference trajectory. The tangent vector
$(\delta q_i, \delta p_i)$ is initialized with random components and normalized to unit Euclidean norm.

At each integration step, the tangent vector is evolved using the linearized equations of motion evaluated along the instantaneous trajectory. The local stretching factor is extracted from the norm growth:
\begin{equation}
g(t) =
\frac{\|\delta \Gamma(t+\Delta t)\|}
{\|\delta \Gamma(t)\|},
\end{equation}
after which the tangent vector is renormalized to unit norm. The largest Lyapunov exponent is then obtained as:
\begin{equation}
\lambda_{\max}
= \frac{1}{T_{\mathrm{meas}}}
\sum_{k} \ln g(t_k),
\end{equation}
where the sum runs over all integration steps after the transient and $T_{\mathrm{meas}}$ denotes the corresponding measurement time.

\subsection*{System sizes and parameter scans}

For the main results presented in the text, we fix the system size to $N=2048$ and compute $\lambda_{\max}$ over $32$ logarithmically spaced energy densities in the range
\begin{equation}
\varepsilon \in [10^{-2}, 10^{2}] .
\end{equation}
For the finite-size scaling analysis shown in the inset of Fig.~1, we consider system sizes
\begin{equation}
N = 32, 64, 128, 256, 512, 1024, 2048,
\end{equation}
and compute $\lambda_{\max}$ at fixed energy densities $\varepsilon = 0.1$, $1.0$, and $10.0$. All finite-size calculations use the same numerical protocol, integration time, and tangent-space renormalization procedure as the main simulations.

\subsection*{Parallelization}

All Lyapunov exponent computations are independent and are therefore parallelized using standard multiprocessing across available CPU cores. No shared state is used between processes, ensuring full reproducibility and consistency between the main results and the finite-size scaling analysis.

\vspace{1cm}
\begin{center}
\rule{0.5\linewidth}{0.4pt}
\end{center}
\end{document}